\documentclass[runningheads,fleqn]{svmult}

\usepackage{makeidx}   
\usepackage{graphicx}  
\usepackage{amssymb}
\usepackage{subeqnar}  
\usepackage{multicol}  
\usepackage{physproc}  
\makeindex             

\begin{document}
%

\title*{What can superconductivity learn from quantized vorticity in $^3$He
superfluids?}
\toctitle{What can superconductivity learn from
quantized vorticity in $^3$He superfluids}
\titlerunning{Unconventional quantized vorticity}

\author{G.~E.~Volovik\inst{1,2}
\and V.~B.~Eltsov\inst{1,3}
\and M.~Krusius\inst{1}}
\authorrunning{ G.~E.~Volovik et al.}
%

\institute{Low Temperature Laboratory, Helsinki University of Technology,
  P.O.Box 2200, FIN-02015
  HUT, Espoo, Finland
\and  Landau Institute for Theoretical Physics, Kosygina 2, 117334
  Moscow, Russia
\and Kapitza Institute for Physical Problems, Kosygina 2, 117334
  Moscow, Russia
}
\date{\today}

\maketitle

\begin{abstract}
In $^3$He superfluids quantized vorticity can take many different
forms: It can appear as distributed periodic textures, as sheets, or
as lines. In the anisotropic $^3$He-A phase in most cases the
amplitude of the order parameter remains constant throughout the
vortex structure and only its orientation changes in space. In the
quasi-isotropic $^3$He-B phase vortex lines have a hard core where the
order parameter has reduced, but finite amplitude. The different
structures have been firmly identified, based on both measurement and
calculation. What parallels can be drawn from this information to the
new unconventional superconductors or Bose-Einstein condensates?
\end{abstract}

\section{Unconventional quantized vorticity}

Soon after the discovery of the $^3$He superfluids in 1972 it was
understood that they represented the first example of unconventional
Cooper pairing among Fermi systems, a p-wave state with total spin
$S=1$ and orbital momentum $L=1$ \cite{Leggett}. This lead to a wide
variety of new phenomena, of which one of the most important is the
discovery of new vortex structures \cite{SalomaaVolovik}. These can be
studied with NMR spectroscopy \cite{JLTP}, when this is combined with
a calculation of the order parameter texture \cite{Karimaki}.

In recent years other unconventional macroscopic quantum systems have been
found and have taken the centre stage. Intermetallic alloys such
as the heavy fermion metals, the high-temperature
superconductors, and the most recent addition, the layered
superconductors of Sr$_2$RuO$_4$ type, do not fit in the
conventional picture of s-wave pairing. Is it possible that
unconventional vortex structures, similar perhaps to some of those
in the $^3$He superfluids, might also be present in these new
systems?

\begin{figure}[t!!!]
\centerline{\includegraphics[width=0.85\textwidth]{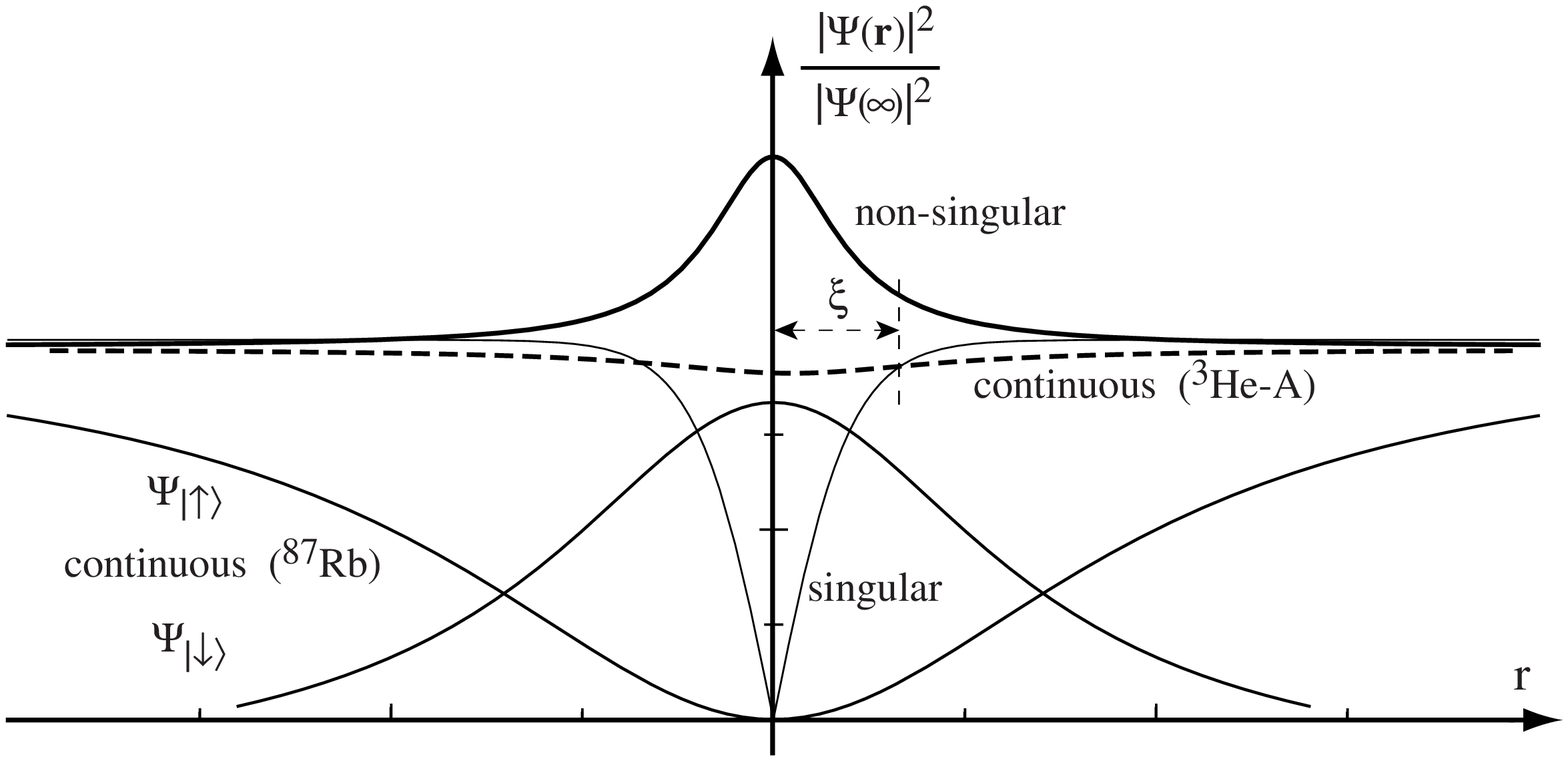}}
\caption[VorCoreSchematics]{Nomenclature of vortex-core structures:
  (i) The {\it singular} vortex has a {\it hard} core whose radius is
  comparable to the superfluid coherence length $\xi(T,P)$ and where the
  order-parameter amplitude vanishes in the center.  (ii) The {\it
    non-singular} vortex has a {\it hard} core in which the order parameter
  has a finite amplitude everywhere within the core. In principle, this
  amplitude can even be larger within the core than far outside. (iii) The
  {\it continuous} vortex has an almost constant order-parameter amplitude
  throughout the {\it soft} core, whose size is much larger than
  $\xi(T,P)$. Within the soft core primarily the orientation of the order
  parameter changes. Examples of {\it continuous} vortices are the doubly
  quantized vortex in $^3$He-A and the vortex in a two-component
  Bose-Einstein condensate, which has a wide inflated core. The latter
  vortex is formed from the condensate fraction $\Psi_{|\uparrow>}$, while
  the soft core is filled by the superfluid component
  $\Psi_{|\downarrow>}$. }
\label{VorCoreSchematics}
\end{figure}

Current belief holds that the superconducting state in the
tetragonal Sr$_2$RuO$_4$ material is described by an order
parameter of the same symmetry class as that in $^3$He-A
\cite{Rice,Ishida}, an anisotropic superfluid with uniaxial
symmetry (where both time reversal symmetry and reflection
symmetry are spontaneously broken).  Recent advances in optical
trapping and cooling of alkali atom clouds to Bose-Einstein
condensates have produced Bose systems which also are described by
a multi-component order parameter: The spinor representation of
the hyperfine spin manifold $F=1$, for instance, would allow the
presence of similar vorticity as in $^3$He-A.

Thus the existence  of unconventional vorticity has moved in the
centre of interdisciplinary debate: To what extent will  reduced
symmetry influence the structure of quantized vorticity? Vortex lines
are defects of the order parameter field, which carry phase winding
and circulation of the respective supercurrent. The conventional
structure is built around a narrow {\it singular hard vortex core}:
The order parameter vanishes in the center of the core. By now it has
been thoroughly verified that quantized vorticity can take many other
forms. Some of the different core structures are schematically listed
in Fig.~\ref{VorCoreSchematics}. There exist differences in
nomenclature in the $^3$He literature and the current theoretical
discussion of vortex line structure in unconventional superconductors,
which we try to unify here.

(i) If the vortex has a {\it hard} core, whose radius is comparable to the
temperature- and pressure-dependent superfluid coherence length $\xi(T,P)$
and where the order-parameter amplitude vanishes in the center, then we
call it {\it singular}. Conventional vortices in the $^4$He-II superfluid
and in s-wave superconductors are singular. Here : $|\Psi({\bf r})|_{{\bf
    r}\rightarrow 0} = 0$. In some approximation such a core can be
pictured to be a tube with a diameter comparable to the coherence length
and filled with normal-state material.

(ii) If the vortex has a hard core but the order parameter has a finite
amplitude everywhere within the core, then we call this a {\it
  non-singular} vortex, as is generally done in the literature on
unconventional superconductivity. Such a core can be viewed as consisting
from a different broken symmetry state than the bulk phase outside the
core. Vortices in $^3$He-B are non-singular: the core of the vortex is made
up of some non-B-phase components of the order parameter, either the axial
(ie. $^3$He-A) or axiplanar state (Sec.~\ref{BrokenSymmetry}). The length
scale, which determines the core radius, is $\xi(T,P)\sim 10$\,--\,100\,nm.

(iii) The {\it continuous} vortex  has an almost constant
order-parameter amplitude throughout the {\it soft} vortex core. By
the  soft core we mean a core whose size is much larger  than
$\xi(T,P)$. Within the soft core primarily the orientation of the
order parameter changes. The larger the soft core diameter, the
smoother is the distribution of the order-parameter amplitude.
Examples of  continuous vortices are the {\it doubly quantized vortex}
in $^3$He-A (Sec.~\ref{ContinuousVortex}) and the vortex in a two-component Bose-Einstein
condensate (Sec.~\ref{SingularToContinuous}). In $^3$He-A the length scale, which determines the
soft-core radius, is the healing length of the dipolar spin-orbit
interaction: $\xi_D(T,P) \sim 10$ -- 40\,$\mu$m.

(iv) A vortex with a {\it composite} core has onion structure: It has a
hard core (with radius $\sim \xi$), which can be either singular or
non-singular, but is embedded within a soft core (with radius $\sim
\xi_D$). (a) The {\it singly quantized vortex} in $^3$He-A is the prime
example (Sec.~\ref{CompositeCore}). It has a hard non-singular core within
a large soft core. The superfluid circulation is generated by the soft
core, but the hard core is needed to satisfy the boundary conditions on the
orbital part of the order-parameter field, such that it becomes continuous
with respect to the bulk fluid. (b) At high enough magnetic field, the
singly quantized vortices in $^3$He-B have composite cores. Here the
vorticity is concentrated in the hard core, while the soft core supports an
inhomogeneous order-parameter distribution where the spin-orbit interaction
is not minimized. This soft core is a deformation in the order-parameter
texture which occurs primarily in the spin part. (c) The {\it spin-mass
  vortex} in $^3$He-B is a composite defect with a narrow hard core, around
which the superfluid circulation is trapped. This core is embedded within a
planar domain-wall-like soliton defect where the spin-orbit interaction is
not minimized and which supports a spin current.

The distinction between non-singular and continuous vortices, and also
the existence of a vortex with a composite core rely upon the presence
of two length scales, which in turn are determined by two energy
scales. In superfluid $^3$He the two relevant energy scales are the
weak spin-orbit interaction and the two orders of magnitude larger
superfluid condensation energy. These fix the soft and the hard core
sizes, respectively, and as a result the soft core is typically two
orders of magnitude larger in diameter than the hard core. If the two
scales become comparable in magnitude (as typically occurs in
superconductors), then the difference between the
continuous and non-singular vortices is washed out.

An understanding of the various structures, in which quantized
vorticity may appear, has led to new insight in the physics of
macroscopic quantum systems. This new understanding now promises to
bring surprising rewards. The discovery of gap nodes in the spectrum
of quasiparticle excitations is generally taken to be a signal for
unconventional pairing in Fermi systems.  An important observation
from recent years is the fact that in the vicinity of these gap nodes
the energy spectrum is linear and the system acquires all the
attributes of relativistic quantum field theory: the analogues of
Lorentz invariance, gauge invariance, general covariance, etc. all are
present.  Therefore fermion superfluids and superconductors  on one
hand and quantum field theory on the other hand show surprising
conceptual similarity. This makes it possible to treat the
condensed-matter quantum systems as laboratory models to study
physical principles which might also be effective in high energy
physics or cosmology. The first examples of such work have been seen
in ``cosmological" laboratory experiments. For instance, it was
recently demonstrated that quantized vortex lines, or linear
topological defects as they are known in field theory, are produced in
quench-cooled transitions from the normal to the
superfluid/superconducting state \cite{BigBang}. This process has been
suggested to mimic the production of cosmic strings in Big-Bang
expansion. A second effort was related to the dynamics of vortex lines
and was exploited to explain the matter-antimatter asymmetry in the
Early Universe \cite{Bevan}, if it is assumed to result from the axial
anomaly of relativistic field theory. Relativistic quantum field
theory may just have found itself an unexpected accomplice!

\section{Special features of $^3$He superfluids}

Superfluid $^3$He has been blessed with the most ideal properties
among the dense coherent quantum systems, approaching those of the
alkali atom clouds in Bose-Einstein condensed states: (i) There are no
bulk impurities since all alien particles are expelled to the
container walls. (ii) The superfluid coherence length $\xi (T,P)$, in
addition to being a function of temperature $T$, also depends on the
applied pressure $P$. By choosing the pressure, the density and
interactions can be varied and $\xi(0,P)$ decreases from 65 nm at zero
pressure to 12 nm at the liquid-solid transition ($P=34.4$ bar). (iii)
For the best wall materials surface roughness can be reduced close to
the level of $\xi (T,P)$. Experimentally this has important
consequences, when the container walls approach ideal solid
boundaries. (iv) Being an isotropic liquid and a Fermi system, liquid
$^3$He is theoretically more tractable than either superconductors or
liquid $^4$He. All complexity and anisotropy is simply derived from
the order parameter, with Cooper pairs in p-wave states \cite{VW}. In
practice this means that experimental observation can be confirmed by
detailed theoretical calculation, although often only retrospectively,
once the true behaviour is already known.

\section{Continuous vortex, skyrmions and merons}
\label{ContinuousVortex}

One of the most striking results emerging from $^3$He research is the
existence of vortices with perfectly continuous singularity-free
structure. In such a vortex the order-parameter amplitude remains
constant throughout the whole structure while its orientation changes
in a continuous manner. In the $^3$He-A phase with axial anisotropy
(and Cooper pairs in $L=S=1$ states), the orbital order-parameter
orientation is denoted by the symmetry axis $\hat{\bf l}$ which points
in the direction of the nodes of the superfluid energy gap. At very
low magnetic fields vorticity is distributed over the entire primitive
cell of the vortex array and thereby takes the form of a periodic
order-parameter texture. At higher fields a region of concentrated
vorticity is formed, the soft vortex core. An integral number of phase
windings is reached along a closed path which encircles the boundary
of the unit cell or the soft core.

\begin{figure}[t!!!]
\centerline{\includegraphics[width=1\textwidth]{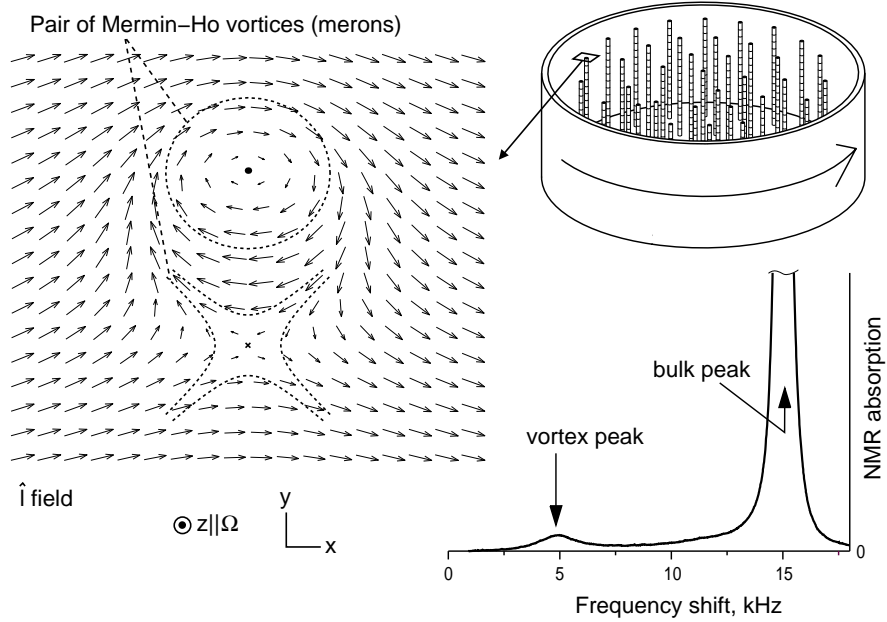}}
\caption[SoftVorCore]{Doubly quantized vortex line in $^3$He-A with
  continuous structure in the soft vortex core: {\it (Top right)} Rotating
  container with
  quantized vortex lines. The pillars depict the soft vortex cores, with a
  diameter of roughly $80\,\mu$m $\gg \xi(T,P)$. Each soft core is
  encircled by a persistent superfluid circulation of two quanta $2\kappa =
  h/m_3 = 0.13$\,mm$^2$/s. {\it (Bottom right)} NMR spectroscopy of
  topological defects in $^3$He-A. In an external magnetic field, which
  exceeds the equivalent of the spin-orbit interaction, $H > H_D \approx
  3\,$mT, the spin $\hat {\bf d}$ and orbital $\hat {\bf l}$ axes are not
  aligned parallel in the soft core. Spin-orbit
  interaction exerts an extra torque on spin precession in nuclear magnetic
  resonance which shifts the NMR frequency \cite{Leggett}. This torque has
  different value within and outside the soft core, and thus gives rise to
  a satellite
  absorption peak. Both the frequency shift and the absorption intensity of
  the satellite are characteristic of the order-parameter texture in the
  soft core \protect\cite{JLTP}. {\it (Left)} Orientational distribution of
  the orbital quantization axis $\hat {\bf l}$ in the soft core,
  depicted in terms of the projection of $\hat {\bf l}$ on the plane
  perpendicular to the vortex axis. The $\hat {\bf l}$ orientations cover a
  solid angle of $4\pi$ and the distribution is everywhere continuous. This
  gives rise to a superfluid circulation of two quanta around the soft core. }
\label{SoftVorCore}
\end{figure}

The simplest possible vortex structure with continuous vorticity in
$^3$He-A is the doubly quantized vortex line. In
Fig.~\ref{SoftVorCore} the orientational distribution of $\hat{\bf l}$
within the soft vortex core is depicted with arrows. By following
their flow, it is noted that all possible $4\pi$ directions of the
radius vector of the unit sphere are present here. This topology of
order-parameter orientations ensures two quanta of circulation
$(N=2)$.

The $4\pi$ topology of $\hat{\bf l}$ orientations within the soft core
is known as a {\it skyrmion}. It can be divided into a pair of {\it
merons} \cite{Meron,Meron2,Meron3} ($\mu\epsilon\rho{\it o}\sigma$ means fraction \cite{Meron}), which in the $^3$He literature are called
Mermin-Ho vortices. In the complete skyrmion the $\hat{\bf l}$ vector
sweeps the whole unit sphere while each meron, or Mermin-Ho vortex,
covers only the orientations in one hemisphere and therefore carries
one quantum of vorticity $(N=1)$.  The meron covering the northern
hemisphere forms a circular $2\pi$ Mermin-Ho vortex, while the
meron covering the southern hemisphere is the hyperbolic $2\pi$
Mermin-Ho vortex.  The centers of the merons correspond to minima in
the potential of trapped spin-wave states, which generate the
satellite peak in the NMR absorption spectrum and make the soft cores
observable in NMR measurement. The satellite from the doubly quantized
vortex was first detected in rotating $^3$He-A in 1982 \cite{Hakonen},
but it is only recently that single-vortex sensitivity was reached and
the quantization number $N=2$ was verified directly
\cite{DoubleQuantumVor}.

Skyrmions and merons are popular structures in physics: For instance,
the double-quantum vortex, in the form of a pair of merons similar to
that in $^3$He-A, is also discussed in the quantum Hall effect where
it is formed by pseudo-spin orientations in the magnetic structure
\cite{Girvin}. In QCD merons are suggested to produce the color
confinement \cite{Meron3}.

In superconductors continuous vortices have been discussed in
Ref.~\cite{BurlachkovKopnin} within a model where  the spin-orbit
coupling between the electronic spins and the crystal lattice is small
and the spin rotation group $SO(3)_S$ is almost exact. The vortex has
essentially the same topology as in $^3$He-A except that instead of
the orbital momentum $\hat{\bf l}$ it is the spin orientations
$\hat{\bf s}$ of the Cooper pairs which cover a solid angle of $4\pi$
in the soft core.

\section{Transformation from singular to continuous vortex}
\label{SingularToContinuous}

From the topological point of view, all vortices with the same winding
number $N$ can be transformed to each other without changing the asymptotic
behaviour of the order parameter, simply by reconstruction of the vortex
core. To obtain the continuous vortex in $^3$He-A (Fig.~\ref{SoftVorCore})
we can start with a $N=2$ {\it singular phase vortex}, which has a hard
core of the size of the coherence length and a uniformly oriented orbital
momentum axis $\hat{\bf l}$ along $\hat{\bf x}$. This is a pure phase
vortex for the orbital component $L_x=+1$ of the order parameter, which
vanishes only on the vortex axis while other components are zero
everywhere. If we now allow for the presence of the other components
$L_x=0$ and $L_x=-1$, then we might fill the hard core with these
components, such that the core becomes non-singular. The non-singular core
can expand further to form a continuous soft core, which is much larger in
radius than the coherence length. Its size is limited by some other weaker
energy scales, for instance in $^3$He-A by the tiny spin-orbit interaction.
Within the core $\hat{\bf l}$ sweeps through all possible orientations and
the topology becomes that of a skyrmion.

This transformation between continuous and singular vorticity has direct
application to Bose-Einstein condensates. Suppose a mixture of two sub
states can be rotated in a laser manipulated trap. Here one starts with a
single Bose-condensate which we denote as the $\mid\uparrow>$
component. Within this component a
pure $N=1$ phase vortex with singular core is created.  Next the hard core
of the vortex is filled
with the second component in the $\mid\downarrow>$ state. As a result the core
expands and a vortex with continuous structure is obtained.  Such a
skyrmion has recently been observed with $^{87}$Rb atoms \cite{Matthews}.
It can be represented in terms of the $\hat{\bf l}$ vector which is
constructed from the components of the order parameter as
\begin{equation}
  \mbox{\hspace*{-5mm}}
\left(
\matrix{
\Psi_\uparrow\cr
\Psi_\downarrow\cr
}\right)=|\Psi_\uparrow(\infty)| \left(
\matrix{ e^{i\phi} \cos {\beta(r)\over 2}
\cr
\sin {\beta(r)\over 2}\cr
}\right)~,~{\hat{\bf l}}=(\sin\beta\cos\phi,-\sin\beta\sin\phi,\cos\beta)
\,\,.
\label{Skyrmion}
\end{equation}
Here $r$ and $\phi$ are the cylindrical coordinates with the axis $\hat{\bf
  z}$ along the vortex axis. The polar angle $\beta(r)$ of the $\hat{\bf l}$
vector changes from 0 at infinity, where $\hat{\bf l} = \hat{\bf z}$ and
only the $\mid\uparrow>$ component is present, to $\beta(0)= \pi$ at the axis,
where $\hat{\bf l} = -\hat{\bf z}$ and only the $\mid\downarrow>$ component is
present (Fig.~1). Thus the vector $\hat{\bf l}$ sweeps the whole unit sphere. As distinct
from $^3$He-A, where the orbital part of the order parameter has three
components and the vortex has $N=2$ circulation, this mixture of
Bose-condensates has two components and the continuous vortex has $N=1$ phase
winding. Various schemes have recently been discussed by which a meron with
phase winding can be created in a Bose-condensate formed within a
three-component $F=1$ manifold \cite{Ho,Nakahara,Marzlin}.

\section{Vortex with composite core}
\label{CompositeCore}

The singly quantized vortex in $^3$He-A has a composite core: It is a
$N=1$ vortex with a non-singular hard core, but where the superfluid
circulation is generated by the soft core.  Thus this $N=1$ vortex is
not a simple $U(1)$ phase vortex, which would have circulation trapped
around a hard core in an otherwise homogeneous $\hat{\bf l}$ texture.
In fact, the function of the hard core is only to provide the
topological stability of the soft core, which with its $2\pi$
orientational distribution of the $\hat{\bf l}$ field produces all the
vorticity. This vortex structure has the lowest energy in a magnetic
field ($H>3\,$mT) at low rotation velocity $(\Omega \lesssim 1$\,rad/s)
\cite{PartsSingular}. Thus it can be created by cooling slowly through
$T_c$ in rotation at low $\Omega$, which secures the equilibrium
vortex state in an adiabatically slow transition. In contrast, if one
starts to rotate the fluid when it is already in the superfluid state
below $T_c$, then the $N=1$ vortex is generally not formed, because
the $N=2$ vortex has lower critical velocity and is created first. The
larger critical velocity of the $N=1$ vortex reflects the much larger
energy barrier involved in the creation of the hard vortex core
(Sec.~\ref{CritVel-CoreSize}). Experimentally the singly and doubly
quantized vortex lines can be distinguished by their very different
NMR absorption satellites.

\section{Vortex sheet}

\subsection{Vortex-sheet structure in $^3$He-A}

\begin{figure}[t!!!]
\centerline{\includegraphics[width=0.9\textwidth]{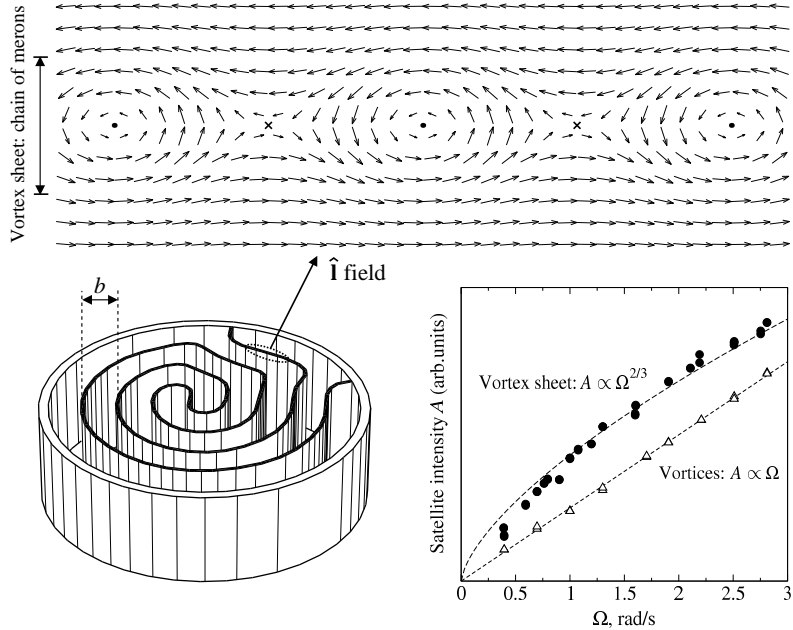}}
\caption[VorSheet]{Vortex sheet in $^3$He-A:
  {\it (Bottom left)} The macroscopic configuration of the equilibrium
  vortex sheet consists of a single continuously folded sheet which fills
  the rotating container evenly. In an axially oriented external magnetic
  field the meander forms a double spiral
  which mimics the original Onsager suggestion of concentric coaxial
  cylindrical surfaces. {\it (Top)} Orbital
  order-parameter texture in the vortex sheet. The arrows depict the
  projection of $\hat {\bf l}$ in the $xy$-plane (perpendicular to the
  rotation axis $\mathbf{\Omega}$). This texture consists of an alternating
  chain of circular (center marked with black dot) and hyperbolic (center
  marked with cross) merons. If one follows a path along the centers of
  merons, the $\hat {\bf l}$ orientation winds continuously: thus the
  vorticity is continuously distributed along the sheet. Outside the sheet
  both $\hat {\bf l}$ and the counterflow velocity ${\bf v} = {\bf v}_n -
  {\bf v}_s$ are oriented parallel to the sheet, but in opposite
  directions on the two sides of the sheet.
  {\it (Bottom right)} The integrated NMR absorption in the
  satellite peak of continuous vortex lines in
  Fig.~\protect\ref{SoftVorCore} follows a linear dependence on the
  rotation velocity $\Omega$ or, equivalently, the number of vortex lines
  in the equilibrium state of rotation. In contrast, if a soliton sheet
  preexists in the container and rotation is slowly increased, then a
  satellite peak with a nonlinear dependence on $\Omega$ is grown.  This is
  the signature of the vortex sheet as a rotating state of superfluid.  }
\label{VorSheet}
\end{figure}

In $^3$He-A the double-quantum vortex line is not the only unconventional
vortex structure with perfect continuity in the order parameter amplitude.
The most unusual continuous structure is the {\it vortex sheet}
\cite{Sheet}, with planar topology (Fig.~\ref{VorSheet}). It consists of a
folded domain-wall-like structure, a soliton sheet, which separates regions
with opposite orientations of the $\hat{\bf l}$ vector. Within this meandering
sheet the continuous Mermin-Ho vortex lines with $N=1$ are confined. These
vortices form a chain of alternating circular and hyperbolic merons, which
are similar to the two constituents of the soft core in
Fig.~\ref{SoftVorCore}. Each meron represents a kink in the soliton
structure and is thus topologically trapped within the soliton: it cannot
exist as an independent object in the bulk liquid outside the soliton. An
analogous example are Bloch lines within a Bloch wall in ferromagnetic
materials. The sheet is attached along two connection lines to the lateral
boundaries. It is through these connection lines that merons with $N=1$
vortex quanta can enter or leave the sheet.

\subsection{Vortex sheet in rotating superfluid}

The vortex sheet is well known from classical turbulence as a thin
interface across which the tangential component of the flow velocity is
discontinuous. Within this sheet vorticity approaches infinity while the
width of the sheet approaches zero \cite{Saffman}.  Historically in
superfluids, vorticity was first suggested to be confined to sheets, when
Onsager \cite{Onsager} and London \cite{London} described the superfluid
state of $^4$He-II under rotation. It soon turned out, however, that in
$^4$He-II a vortex sheet is unstable towards break-up into separated
quantized vortex lines. Nevertheless, a calculation on the sheet spacing by
Landau and Lifshitz \cite{Landau}, who did not impose a quantization
requirement, happens to be exactly to the point for the vortex sheet in
$^3$He-A. Here the vortex sheet proved to be stable owing to the
topological confinement of the vorticity within the topologically stable
soliton sheet \cite{Sheet}.

The equilibrium state of the vortex sheet in a rotating vessel is
constructed by considering the kinetic energy from the flow between the
folds and the surface tension $\sigma$ from the soliton sheet. It is then
concluded that the distance between the parallel folds has to be $b = (3
\sigma/ \rho_{s \parallel})^{1/3} \; \Omega^{-2/3}$. This is somewhat
larger than the inter-vortex distance in a cluster of vortex lines. The
areal density of circulation quanta has approximately the solid-body value
$n_v = 2\Omega/\kappa$. This means that the length of the vortex sheet per
two circulation quanta is $p = \kappa/(b\Omega)$, which is the periodicity
of the order-parameter structure in Fig.~\ref{VorSheet} ($p\approx
180\,\mu$m at $\Omega = 1\,$rad/s).  The NMR
absorption in the vortex-sheet satellite is proportional to the total
volume of the sheet which in turn is proportional to $1/b \propto
\Omega^{2/3}$. This nonlinear dependence of the satellite absorption on
rotation velocity is one of the experimental signatures.  Locally the
equilibrium vortex sheet corresponds to a configuration with accurately
equidistant layers. This is manifested by Bragg reflections of spin waves
between the folds of the sheet, which produces a characteristic frequency
shifted absorption in the observed NMR line shape \cite{Bragg}.

\subsection{Vortex sheet in superconductor}

The vortex sheet has also been discussed in unconventional superconductors
\cite{Sigrist1989,Sigrist98}. Similar to superfluid $^3$He-A the vorticity
is trapped in a domain wall which separates two domains with opposite
orientations of the $\hat{\bf l}$-vector \cite{VolovikGorkov1985}.  But,
unlike the case of $^3$He-A, the trapped kink is not a meron but a singular
vortex with the fractional winding number $N=1/2$ (in contrast to isolated
vortex lines which are singly quantized). If there are many trapped
fractional vortices, then they form a vortex sheet, which as suggested in
Ref.~\cite{Sigrist98}, can be responsible for the flux flow dynamics in the
low-temperature phase of the heavy-fermion superconductor UPt$_3$. In
superconductors a vortex-sheet-like structure could also appear
dynamically, when it is topologically not stable. This state would
correspond to a smectic phase of the flowing vortex matter, as was argued
in the case of eg.  NbSe$_2$ \cite{SmecticVortexLattice}.

\section{Fractional vorticity and fractional flux}

An unusual object is the {\it half-quantum vortex}. Here the
order-parameter phase changes by $\pi$ on circling once around this
line. The change of sign of the order parameter can be compensated by
some extra degree of freedom, which usually is the spin. Such a linear
structure was originally predicted to appear in $^3$He-A
\cite{VolMin}, but has not yet been observed there experimentally.
Later it was also predicted to appear in unconventional superconductors
\cite{Geshkenbein1987}. Some years ago it was discovered as the
intersection line of three grain boundary planes in a thin film of
YBa$_2$Cu$_3$O$_{7-\delta}$ \cite{Kirtley1996}. The half-quantum
vortex has also been suggested to exist in Bose-Einstein condensates
with a hyperfine spin $F=1$ \cite{Leonhardt}.

Based on the $^3$He-A example more possible structures of the
fractional vortices in unconvetional superconductors can be predicted.
In $^3$He-A the discrete symmetry, which supports the half-quantum
vortex, arises from the changes in sign when the spin axis $\hat{\bf
d}$ and the orbital axis $\hat{\bf e}_1+i\hat{\bf e}_2$ are taken once
around the line and rotate into $-\hat{\bf d}$ and  $-(\hat{\bf
e}_1+i\hat{\bf e}_2)$. When both of these axes change sign, then the
order parameter returns to its initial value
\begin{equation}
\hat {\bf d}=\hat {\bf x}\cos {\phi\over 2} +\hat {\bf y}\sin {\phi\over
2}~~,~~
 \hat{\bf e}_1+i\hat{\bf e}_2=(\hat {\bf x}+i\hat {\bf y})e^{i\phi/2}~.
\label{HalfQuantumVortex}
\end{equation}
Here $\phi$ is the azimuthal angle of the cylindrical coordinate system and
the magnetic field is applied along $\hat {\bf z}$ to keep $\hat{\bf d}$ in
the $xy$ plane. The spin axis $\hat{\bf d}$ rotates by $\pi$ on circling
around the half-quantum vortex. Thus a ``spectator'' in $^3$He-A, who
travels around the vortex, would find its spin reversed with respect to the
spin of a ``spectator'' who was at rest.  This situation is the analog of
the Alice string in particle physics \cite{Schwarz} where a particle, which
moves around the string, in a continuous manner flips its charge or parity
or enters the ``shadow'' world \cite{Silagadze}.

In superconductors the crystalline structure must be taken into
account. In the simplest representation which preserves tetragonal
symmetry, the $p$-wave order parameter in  Sr$_2$RuO$_4$ has the form
\begin{equation}
\Delta({\bf k})=\Delta_0~(\hat{\bf d}\cdot{\mathbf \sigma})\left(\sin
{\bf k}\cdot{\bf a}  + i~\sin {\bf k}\cdot{\bf b} \right)e^{i\theta}~,
\label{ChiralOP}
\end{equation}
where ${\bf k}$ is momentum, $\theta$ is the phase of the order parameter, ${\bf a}$ and
${\bf b}$ are the elementary vectors in the basal plane of the crystal
lattice. Vortices with fractional quantization $N$ can now be
constructed in two ways. If the $\hat {\bf d}$-field is sufficiently
flexible, the analog of the vortex with $N=1/2$ in
Eq.~(\ref{HalfQuantumVortex}) becomes possible, where $\hat {\bf d}
\rightarrow -\hat {\bf d} $ and $\theta \rightarrow \theta + \pi$ on
circling once around the vortex \cite{Maki}. Another possibility is
the M\"obius-strip geometry. Here the crystal axes ${\bf a}$ and ${\bf
b}$ are twisted continuously  by the angle $\pi/2$ on traversing
around the closed wire loop \cite{Monopoles}. This closed loop traps
fractional flux, since the local orientation of the crystal lattice
continuously changes by $\pi/2$ around the loop, ${\bf a} \rightarrow
{\bf b} $ and ${\bf b} \rightarrow -{\bf a}$, which means that the
order parameter becomes multiplied by $i$. The single-valuedness of
the order parameter requires that this change must be compensated by a
change in the phase $\theta$ by $\pi/2$. As a result the phase winding
around the loop is $\pi/2$ and  $N=1/4$.

This, however, does not mean that such a M\"obius loop  in a chiral
$p$-wave superconductor traps 1/4 of the magnetic flux $\Phi_0$ of a
conventional Abrikosov vortex. Because of the breaking of time
reversal symmetry in chiral crystalline superconductors, persistent
electric currents arise not only due to phase coherence but also due
to deformations of the crystal \cite{VolovikGorkov1984}:
\begin{equation}  {\bf j}=\rho_s\left( {\bf v}_s-{e\over mc}
{\bf A}\right) + Ka_i\nabla b_i~,~{\bf v}_s={\hbar \over 2m}\nabla \theta ~.
\label{SuperfluidCurrent}
\end{equation}
The magnetic flux trapped in the loop is obtained from the condition
of zero current,  ${\bf j}=0$ in Eq.~(\ref{SuperfluidCurrent}). Thus,
the trapped flux depends on the parameter $K$ in the deformation
current. In the limiting case of $K=0$ the flux is $\Phi_0/4$ (or
$\Phi_0/6$ if the underlying crystal lattice has hexagonal symmetry).

In a nonchiral $d$-wave superconductor of layered cuprate-oxide
structure the order parameter can be represented by:
\begin{equation}
\Delta({\bf k})=\Delta_0~\left(\sin ^2{\bf k}\cdot{\bf a} -\sin^2{\bf
k}\cdot{\bf b} \right)e^{i\theta}~.
\label{DWaveOP}
\end{equation}
The same twisted wire loop which transforms ${\bf a} \rightarrow
{\bf b} $ and ${\bf b} \rightarrow -{\bf a}$, produces a change of
sign of the order parameter, which must be compensated by a change of
the phase $\theta$ by $\pi$. This corresponds to a circulation of half
a quantum $N=1/2$, i.e. the fractional flux trapped by this loop is
$\Phi_0/2$, since the parameter $K$ in Eq.~(\ref{SuperfluidCurrent})
is exactly zero in nonchiral superconductors.  The same reasoning
gives rise to the $\Phi_0/2$ flux associated with the intersection
line of three grain boundary planes, the tri-crystal line
\cite{Kirtley1996}: around this line ${\bf a} \rightarrow {\bf b} $
and ${\bf b} \rightarrow -{\bf a}$. Only the observation of a
fractional flux different from $\Phi_0/2$ would indicate the breaking
of time reversal symmetry
\cite{Sigrist1989,VolovikGorkov1984,Sigrist1995}.

\section{Broken symmetry in the vortex core}
\label{BrokenSymmetry}

In the quasi-isotropic $^3$He-B phase (with Cooper pairs in the total
angular momentum state $J=0$), the simplest possible vortex is a
singular vortex with $N=1$, where all the order-parameter components
are zero on the axis of the vortex core. However, this vortex is never
realized. All vortices (and other linear defects) observed in $^3$He-B
have a non-singular hard core. This is a typical situation for
superfluids/superconductors with a multi-component order parameter. As
a rule the superfluid/superconductor does not tolerate a full
suppression of the superfluid fraction in the core, if there is a
possibility to escape this by filling the core with other components
of the order parameter. This rule applies also to high-temperature
superconductors with $d$-wave pairing, where the $N=1$ vortex supports
a nonzero $s$-wave component in the vortex core
\cite{VolovikVortex}.

\begin{figure}[t!!!]
\centerline{\includegraphics[width=1\textwidth]{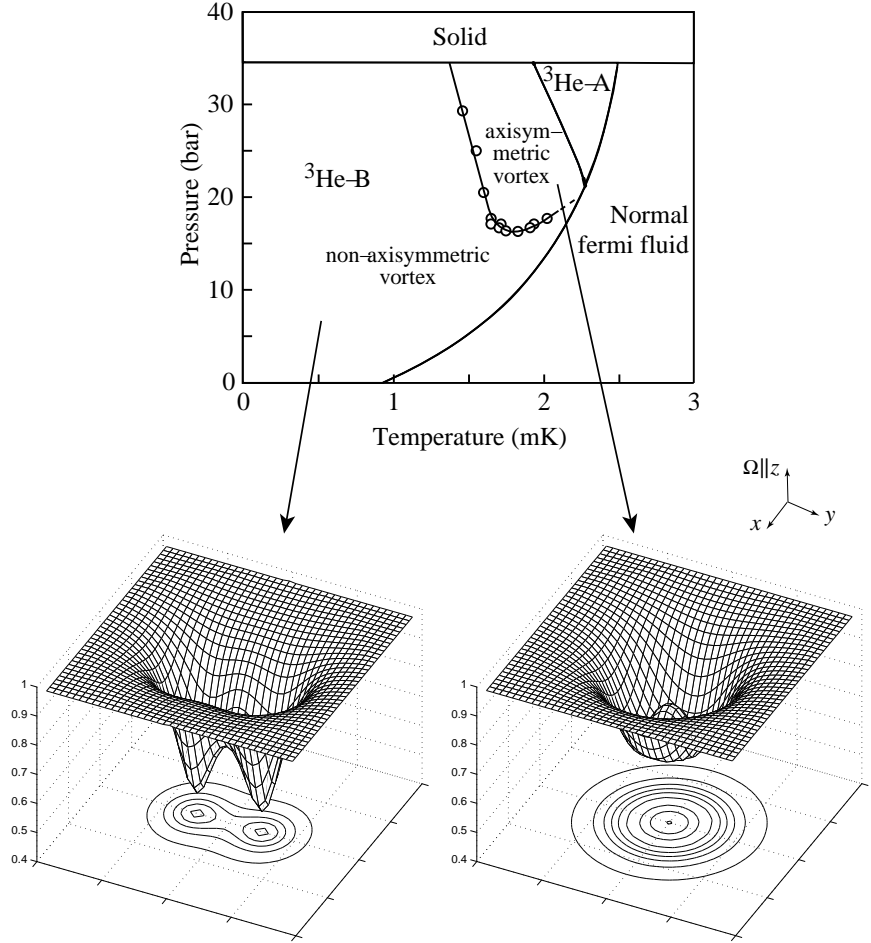}}
\caption[VorCoreTransition]{Phase transition in the structure of the
singular vortex core in $^3$He-B: {\it (Top)} Pressure vs. temperature
phase diagram of $^3$He liquids, with the phase transition line for
the B-phase vortex core structure. The exact intercept of the
transition line at $T_c$ is unknown. {\it (Bottom)} Both measurement
\protect\cite{FerromagneticCoreExp} and calculation
\protect\cite{CoreTransition} confirm that the axi-symmetric core at
high temperatures spontaneously reduces its symmetry and transforms to
a double-well structure at low temperatures. The magnitude of the
square of the order-parameter amplitude has here been plotted in the
$xy$-plane (perpendicular to the rotation axis) for the two core
structures. This quantity remains finite throughout the cross section
of the core. } \label{VorCoreTransition}
\end{figure}

\subsection{Vortex core transition}

The first NMR measurement on rotating $^3$He-B revealed as a function
of temperature and pressure a first order phase transition in the
vortex core structure \cite{Ikkala}. This phenomenon in
Fig.~\ref{VorCoreTransition} was the first example of a phase
transition in the structure of any topological defect. Other
transitions have been identified in $^3$He-A since then, but the
B-phase transition remains the most prominent one.  This transition
separates two vortices with the same topology ($N=1$), but with a
different structure of the hard core. The existence of the transition
illustrates that vortices in $^3$He-B have to be non-singular and have
a complex structure of the hard core. Later these two stable vortex
core structures  were theoretically identified as two different
stationary minimum energy solutions as a function of pressure  in the
Ginzburg-Landau temperature regime. Both structures have broken parity
in the core.

\subsection{Ferromagnetic core}

The high-pressure $^3$He-B vortex has an axi-symmetric core in which
the A-phase order-parameter components dominate. These components produce
increase in the order parameter amplitude close to vortex axis which is
shown in Fig.~\ref{VorCoreTransition}, bottom right. This non-singular
vortex with a superfluid core lies lower in energy, than the simplest
most symmetric solution with a normal core, and displays ferromagnetic
spin polarization \cite{FerromagneticCoreTheory}, which is observed in
the rotating NMR experiments as a gyromagnetism
\cite{FerromagneticCoreExp}.

Similar, but antiferromagnetic cores have been discussed for
high-$T_c$ vortices within the popular $SO(5)$ model for
superconductivity and antiferromagnetism \cite{SO5}. Here it has been
established that in the Ginzburg-Landau regime, in certain regions of
the parameter values, a solution with normal vortex cores is unstable
with respect to non-singular antiferromagnetic cores \cite{SO5vortex}.
Also non-singular vortex cores in the heavy fermion superconductor
UPt$_3$ could possibly explain why a three times larger flux-flow
resistivity is observed parallel to the $\hat c$ axis compared to the
perpendicular directions \cite{EnhancedCore}.

\subsection{Asymmetric double core}

The low-temperature B-phase vortex has a non-axisymmetric core, i.e.
the axial $U(1)$ symmetry of the ferromagnetic core is spontaneously
broken, to create a dumbbell-like double core
(Fig.~\ref{VorCoreTransition}, bottom left). It can be considered as
a pair of half-quantum vortices, connected by a non-topological wall
\cite{SalomaaVolovik,CoreTransition,SalomaaVolovik1989}. The
separation of the half-quantum vortices increases with decreasing
pressure and thus the two-core structure is most pronounced at zero
pressure \cite{Volovik1990}. The asymmetry of the core has been
verified by direct observation of a Goldstone mode which is a direct
consequence from the loss of axial symmetry: the deformed vortex core
can become twisted in the presence of a special type of B-phase NMR
mode, which is then detected as a reduction in rf absorption
\cite{Kondo1991}.

Related phenomena are also possible in superconductors. In
Ref.~\cite{Zhitomirsky} the splitting of the vortex core into a pair
of half-quantum vortices has been discussed in heavy-fermion
superconductors. In fact a vortex-core splitting may have been
observed in high-$T_c$ superconductors \cite{Hoogenboom}. These
observations were interpreted as tunneling of a vortex between two
neighboring sites in the potential wells created by impurities.
However, the phenomenon can also be explained in terms of vortex-core
splitting.

\section{Vortex formation by intrinsic mechanisms}

In addition to vortex structure, the second most important question
becomes the creation of the different forms of vorticity. Vortex
formation by intrinsic mechanisms is a topic which has been discussed
for decades in superfluidity, but which has been notoriously
difficult. The problems are caused by interference from extrinsic
effects, primarily from remanent vorticity trapped on rough surfaces.
One of the major developments from the last ten years has been the
emergence of reliable measurements on critical velocities in
$^4$He-II. These have been performed by monitoring the superflow
through sufficiently small sub-micron-size orifices
\cite{AvenelVaroquaux,PackardRMP}. Intrinsic vortex formation has thereby become
an important phenomenon in superfluids -- quite unlike in
superconductors, where typically vortices appear due to different types of
extrinsic effects at the lower critical field $H_{c1}$. However, it is
useful even in the case of superconductors to keep in mind the more
ideal properties of vortex formation. Also the process of unpinning of
vortices which is important for a problem of vortex creep in
superconductors can be discussed in terms of vortex nucleation: The motion
of vortex from the pinning site is equivalent to creation of a vortex loop
which annihilates the pinned part of the vortex line.

\subsection{Nucleation barrier}

Many vortex phenomena, not only nucleation, but also the unpinning of
remanent vorticity, involve energy barriers which are usually overcome
by thermal activation. At the lowest temperatures quantum tunneling
has been suggested as a possible mechanism, where a macroscopic amount
of matter is assumed to participate coherently in a tunneling process.
Vortex nucleation in orifice-flow of $^4$He-II displays a
characteristic low-temperature plateau in the temperature dependence
of the critical velocity which has been claimed to support the quantum
tunneling concept \cite{QuantumTunnel}. This question is also
discussed in superconductors in the context of unpinning and creep.
However, firm proof for such interpretation is still missing.

In $^3$He-B the critical velocity  was measured in rotating
experiments with single-vortex resolution in the early 1990's. This
proved to be a more straightforward measurement than in $^4$He-II. The
measured temperature dependence of the critical velocity resembles
that of the superfluid energy gap $\Delta (T,P)$
\cite{SingleVortexNucleation}, which  at the lowest temperatures also
approaches a temperature-independent plateau. The explanation here,
however, does not involve quantum tunneling, but the superflow
instability. This phenomenon resembles a second order transition where
the energy barrier goes to zero as a function of the scanned variable,
in this case the superflow velocity.

When a cylinder with superfluid $^3$He-B is slowly accelerated to
rotation, the state with one single vortex line becomes energetically
favorable when the superflow at the circumference exceeds the Feynman
critical velocity $v_{c1} = \kappa /(2\pi R) \ln{(R/r_c)}$
\cite{Donnelly}. With a container radius $R$ of a few millimeters and
a circulation quantum $\kappa = h/(2m_3) = 0.066$ mm$^2$/s, this
velocity is only $10^{-2}$ mm/s. Above this velocity remanent
vorticity, which has been trapped on the cylinder wall, may start to
unpin and then to expand to rectilinear vortex lines. It is the
equivalent of $H_{c1}$ in superconductors. However, if we exclude
extrinsic mechanisms of vortex formation, then the vortex-free state
will persist metastably to much higher velocities because of the
nucleation barrier.

In $^3$He-B the nucleation barrier is practically impenetrable. The
argument is the following: The vortex is nucleated on the wall as a
segment of a vortex ring. The radius of a ring sustained by superflow
at the velocity $v_s$ can be written as $ r_\circ = (\kappa/4\pi v_s)
\ln {(r_\circ/r_c)} $, where the vortex-core radius $r_c$ is of the
order of the superfluid coherence length $\xi_{3_{\rm He}} \sim 10$ --
100 nm. The energy of a ring is $E(v_s) = {1 \over 2} \rho_s \kappa^2
r_\circ \ln {(r_\circ/r_c)}$, where $\rho_s \sim m/a^3$ is the
superfluid density, and $a$ the interatomic distance. This energy
constitutes the nucleation barrier. On dimensional grounds we may
write $E(v_s)/k_B T \sim (r_\circ/a) (T_F/T) \ln{(r_\circ/r_c)}$,
where $T_F = \hbar^2/2ma^2 k_B \sim 1\,$K is the degeneracy
temperature of the $^3$He quantum fluid. Assuming that $r_\circ >
r_c$, we find $E(v_s)/k_B T > 10^5 \ln{(r_\circ/r_c)}$. Such a barrier
height in $^3$He-B is inaccessible at all temperatures below $T_c$. In
contrast in $^4$He-II,  $\xi_{4_{\rm He}} \sim a$, and the barrier is
low, $E(v_s)/k_B T > \ln{(r_\circ/r_c)}$. It can be thermally
overcome, except at the lowest temperatures below 0.2\,K.

\subsection{Vortex formation in a hydrodynamic instability}

The huge barrier in  $^3$He-B means that the vortex formation
mechanism cannot be thermal activation. When the superflow
velocity is increased (by increasing the rotation) a threshold
$v_{cb}$ is finally reached above which homogeneous flow loses
local stability. This occurs when the energy density of the
superflow, $\rho_s v_s^2/2$, exceeds the energy responsible for
the topological stability of a vortex, which is of order
$\rho_s(\kappa/2\pi r_c)^2$. An order of magnitude estimate of the
maximum velocity is thus $v_{cb}\sim {\kappa /2\pi r_c}$. At this
velocity, the radius $r_\circ$ of the nascent ring becomes
comparable to $r_c$ and the nucleation barrier goes to zero. The
instability inevitably leads to the creation of a vortex when no
other mechanism is available at lower $v_s$.

\begin{figure}[t!!!]
\centerline{\includegraphics[width=0.85\textwidth]{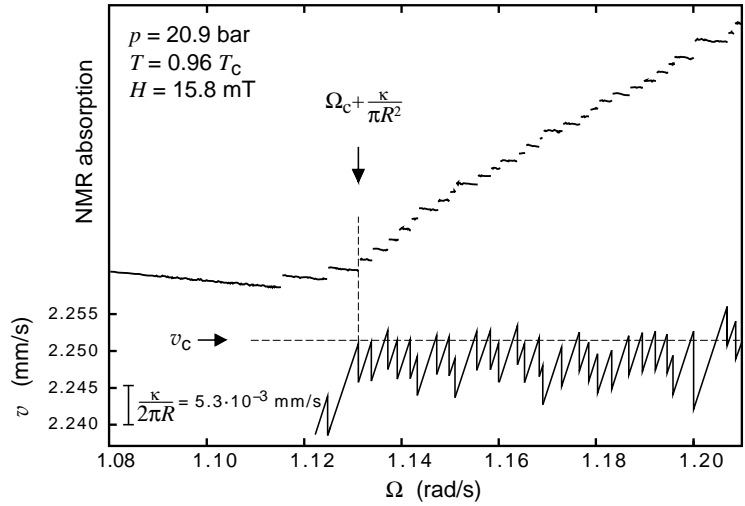}}
\caption[CritVelMeasurement]{ Single-vortex formation as a function of
rotation velocity $\Omega (t)$, for $^3$He-B in a cylindrical
container. {\it Top:} The vertical axis denotes the height of a NMR
absorption peak, where the intensity increases per each newly added
vortex line by a constant amount. Vortex formation starts with a first
step increase at $\Omega_c = 1.115\,$rad/s, but the critical threshold
is identified from the third step (dashed vertical line) where the
maximum flow velocity $v_c$ reaches a stable value (dashed horizontal
line in the plot at bottom). {\it Bottom:} The corresponding superflow
velocity $v_s(R)$ at the cylinder wall. Each step increase in the
upper plot corresponds to a drop by $\kappa/(2\pi R)$ in the velocity
$v_s(R)$, the equivalent of one circulation quantum $\kappa = h/(2m_3)
= 0.066\,$mm$^2$/s. The average of the maximum velocities $v_s$
defines the critical velocity $v_c$. The scatter from the average is
probably of experimental origin. The rotation is here increased at a
constant slow rate ($d\Omega / dt = 2 \cdot 10^{-4}$ rad/s$^2$), the
sample is contained in an epoxy-resin-walled cylinder with radius
$R=2\,$mm at a pressure $P=20.9$ bar, in a magnetic field
$H=15.8\,$mT, and temperature $T=0.96\,T_c$. (From
Ref.~\protect\cite{SingleVortexNucleation}.) }
\label{CritVelMeasurement}
\end{figure}

\begin{figure}[t!!!]
\centerline{\includegraphics[width=0.8\textwidth]{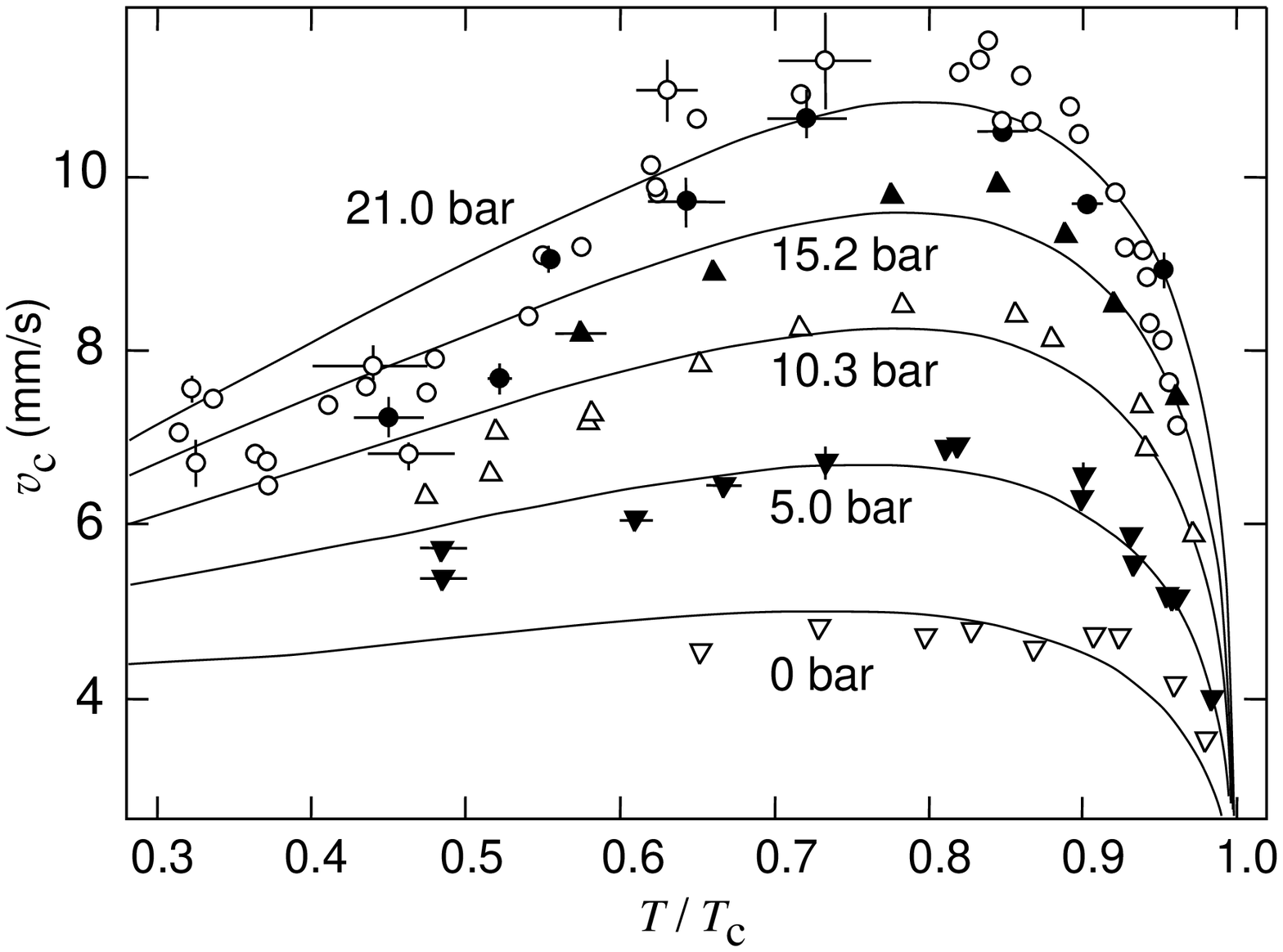}}
\caption[CritVelTempDependence]{ Measurement of critical flow velocity
  $v_c$ in $^3$He-B {\it vs.} temperature for pressures $P \leq 21$ bar.
  The solid curves have been fitted to the data with $v_c = v_{cb}
  (\xi/d)^{\chi}$, where the fitting parameters $d$ and $\chi$ characterize
  the surface roughness on the cylinder wall, $v_{cb}(T,P)$ is the
  calculated bulk liquid superflow instability \protect\cite{VMS} and
  $\xi(T,P)= \xi(0,P) [\Delta(0)/\Delta(T)] $ the superfluid coherence
  length. The roughness is modeled by the protuberance with height $d$ and
  apex angle $\pi/(1-\chi)$ which acts as the nucleation center. The fit
  gives $d = 3.1\,\mu$m and $\chi =0.45$ (apex angle $\approx 30^{\circ}$).
  The measurements have been performed with an epoxy-resin-walled container
  with radius $R=3.5\,$mm. (From
  Ref.~\protect\cite{SingleVortexNucleation}.) }
\label{CritVelTempDependence}
\end{figure}

In $^4$He-II well below $T_\lambda$, the estimate of $v_{cb}$ in the
form $\kappa /2\pi \xi_{4_{\rm He}}$ agrees in order of magnitude with
the Landau velocity, defined by the roton gap $\Delta_r$ and momentum
$p_r$ as $v_L = \Delta _r /p_r \sim 60$ m/s \cite{Donnelly}. However,
this limiting velocity is not observed directly in the measurements of
orifice flow, by extrapolating the thermally activated process to $T
\rightarrow 0$. The reason is that the measured quantity is the
average flow velocity through the aperture and not the local critical
velocity at the nucleation site. On the circumference of the orifice,
the local velocity will be enhanced from the average value by surface
roughness, in particular when the superflow is deflected around any
sharp protuberances which match the length scale $r_\circ$ of the
evolving vortex half ring. The most effective of such excrescences on
the circumference will then selectively become the nucleation center
\cite{SingleVortexNucleation}. For this reason the measured critical
velocity is expected to be roughly a factor of $\lesssim 10$ smaller
than the ideal limiting value (Fig.~\ref{CritVelMeasurement}).

In $^3$He-B, the estimate of $v_{cb} \sim {\kappa /2\pi \xi_{3_{\rm He}}}$ is
smaller by 3 orders of magnitude than in $^4$He-II, but again
comparable to the appropriate Landau limit, defined by the energy
gap and the Fermi momentum as $\Delta(T) /p_F$. In this case the
velocity $v_{cb}$ is also known from direct calculations of the
stability limit of homogeneous superflow \cite{VMS}. As in orifice
flow, the measured average velocity at vortex formation is smaller
than the calculated bulk $v_{cb}$ and depends on surface
roughness. The principal difference from $^4$He-II is the much
larger length scale $r_\circ \sim \xi_{3_{\rm He}}$, which means that
experimentally the influence of surface roughness is less
prominent, remanent vorticity can be avoided in the presence of
sufficiently smooth surfaces, and stable periodic vortex formation
can be investigated with bulk liquid flowing past a flat wall.
This is in stark contrast to $^4$He-II, where the coherence length
is of atomic size, surface roughness generally provides an
unlimited source of trapped remanent vorticity, and intrinsic
nucleation can only be observed in flow through a sufficiently
small orifice.

\begin{figure}[t!!!]
\centerline{\includegraphics[width=0.85\textwidth]{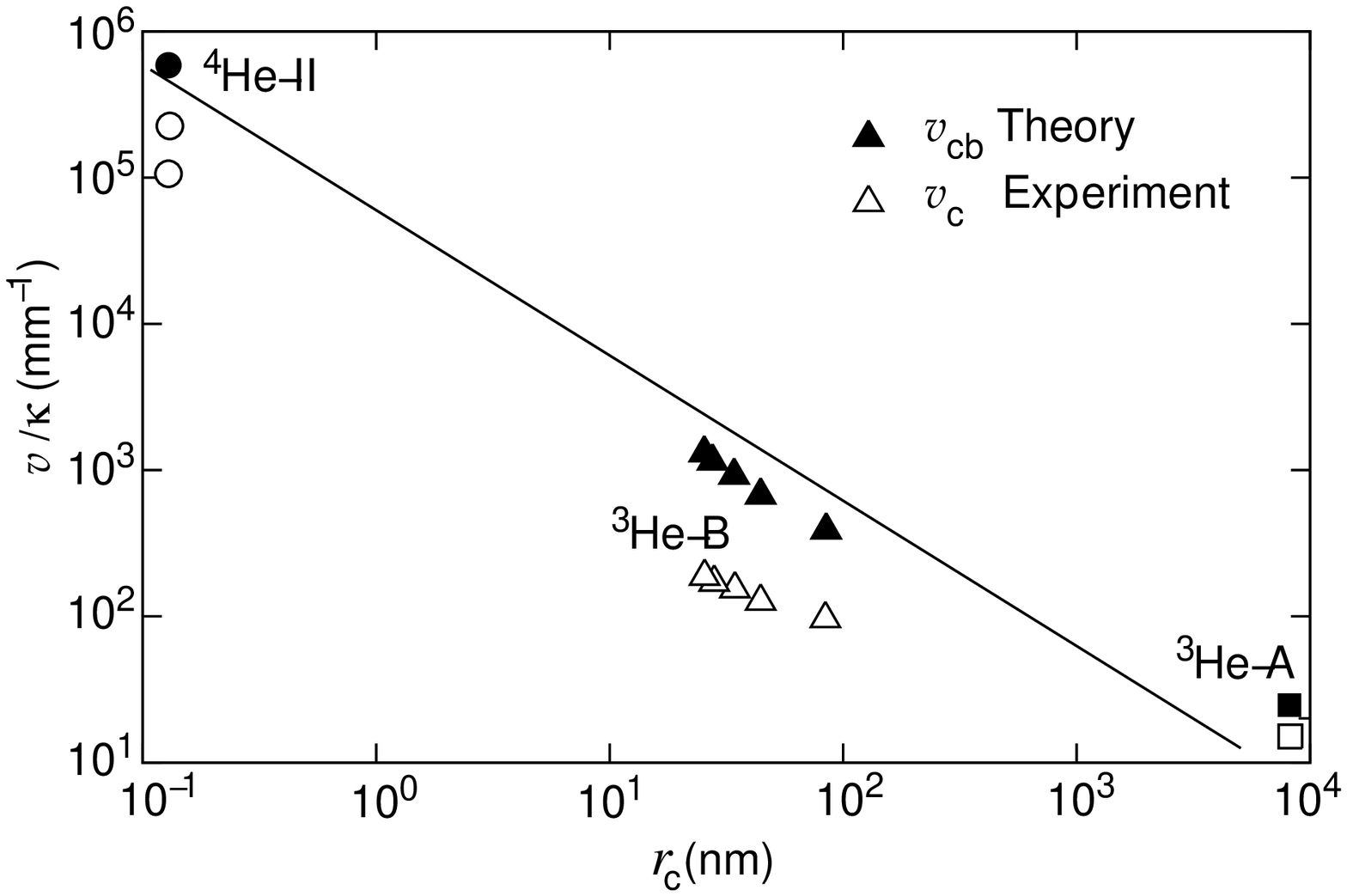}}
\caption[CritVelLengthScaleDependence]{ Theoretical
  $v_{cb}/\kappa$ and maximum experimental $v_{c,max}/\kappa$ plotted for 3
  superfluids as a function of their core size $r_c$.  For $r_c$ we use 0.1
  nm in $^4$He-II, in $^3$He-B it is the superfluid coherence length
  $\xi(T,P)$, and for continuous vortices in $^3$He-A in a magnetic field
  the spin-orbit healing length $\xi_D$. In $^4$He-II we use the Landau
  limit $v_L$ for $v_{cb}$, in $^3$He-B $v_{cb}$ is the calculated maximum
  superflow velocity \protect \cite{VMS}, and in $^3$He-A it corresponds to
  the helical textural instability \protect \cite{RuutuA,Kopu}. For
  $v_{c,max}$ in $^4$He-II the measured value in Ref.  \protect
  \cite{QuantumTunnel} is used, for $^3$He-B data from
  Ref.~\protect\cite{SingleVortexNucleation}, and for $^3$He-A from
  Ref.~\protect \cite{RuutuA}. The line is a guide for the eye, but it
  obeys the relation $v_c/\kappa \propto 1/r_c$. (From
  Ref.~\protect\cite{SingleVortexNucleation}) }
\label{CritVelLengthScaleDependence}
\end{figure}

Owing to the excessively high nucleation barrier, in $^3$He-B it
must be the velocity $v_{cb}$ of superflow instability which becomes the
appropriate velocity of vortex formation. The process then
corresponds to a classical instability, which occurs at the
pair-breaking velocity. Thus in the case of $^3$He-B, the reason for a
plateau in $v_c(T)$ in the $T \rightarrow 0$ limit is that the characteristic
physical quantities, such as the gap amplitude $\Delta (T)$, which
determine the instability velocity, become temperature
independent. Consequently not only quantum tunneling, but also the
intrinsic instability provides an explanation for the
low-temperature plateaus which are observed in many different
systems, including the case of $^4$He-II at the lowest
temperatures.

\subsection{Formation of continuous vortex lines: Dependence of critical
velocity on core size}\label{CritVel-CoreSize}

The order-of-magnitude expectation for the bulk superflow instability
is $v_c \sim \kappa/(2\pi r_c)$, where $r_c$ is the core size of the
emerging vortex. For $^4$He-II and $^3$He-B it is the size of the hard
core, which is on the order of the superfluid coherence length. For
the continuous vortex in $^3$He-A the length scale of the  soft core
structure is much larger: It is the healing length $\xi_D \gtrsim
10\,\mu$m of the spin-orbit coupling (Fig.~\ref{SoftVorCore}). The
same length scale also applies to the structure of the soliton and
vortex sheets. Because of this long length scale the measured critical
velocities in the A phase are two orders of magnitude lower than in
the B phase, as shown in Fig.~\ref{CritVelLengthScaleDependence}. This
explains why the continuous $N=2$ vortex is formed in the A liquid
when it is accelerated into rotation, rather than the composite $N=1$
vortex with a hard core, although the latter might be energetically
preferred.

\begin{figure}[t!!!]
\centerline{\includegraphics[width=0.85\textwidth]{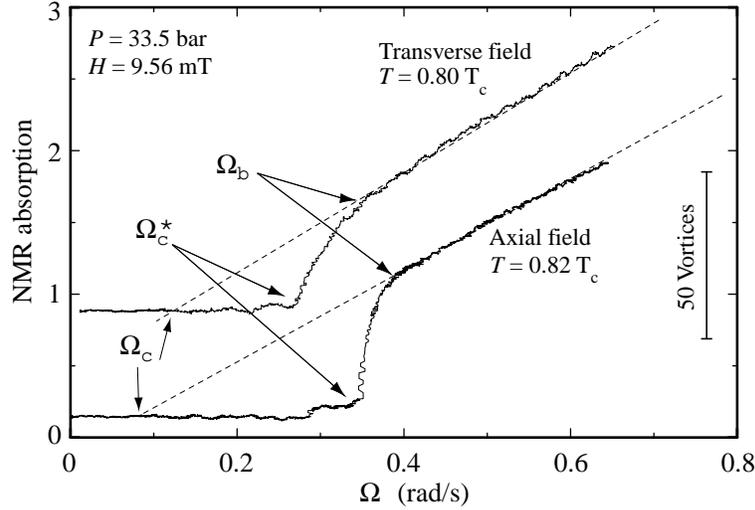}}
\caption[TextureTransition]{ Two examples of vortex-line formation in
  a transition of the order-parameter texture in $^3$He-A as a function of
  the applied rotation $\Omega$. The texture transition occurs at
  $\Omega_c^*$, where a large number of vortex lines is simultaneously
  formed. After the transition in the transformed texture the regular
  periodic vortex formation process sets in, where the flow is limited by a
  constant critical velocity $v_c = \Omega_c R$. This process gives a
  linear dependence for the number of vortex lines $\cal N$ as a function
  of $\Omega$, with the slope $d{\cal N}/d\Omega = 2\pi R^2/\kappa$.  }
\label{TextureTransition}
\end{figure}

At the container wall the boundary condition on the order parameter
requires that the orbital quantization axis $\hat{\bf l}$ has to be
oriented along the surface normal. Therefore the center of vortex
formation must evolve outside a surface layer with a width $\sim
\xi_D$, out of the influence of surface roughness. This has been
recently experimentally confirmed \cite{DoubleQuantumVor}. Here the
only mechanism available is an instability of the order-parameter
texture.

An example of a critical process, which exemplifies the fact that the
order-parameter texture is involved, is shown in
Fig.~\ref{TextureTransition}. Here the vertical axis is the peak
height of the satellite absorption in Fig.~\ref{SoftVorCore}, which is
proportional to the number of vortex lines, and the horizontal axis is
the external drive. Initially on increasing $\Omega$, no vortex lines
are formed. When the critical velocity $\Omega_c^*$ is reached, a
large number of vortex lines is suddenly simultaneously formed. During
further increase of $\Omega$ the system recovers and a characteristic
linear slope is retrieved. The linear dependence represents a
reproducible periodic process where one vortex at a time is formed at
a constant critical velocity, similar to that for B-phase vortices in
Fig.~\ref{CritVelMeasurement}. The extrapolation of the linear
dependence back to zero vortex number gives for the onset a
lower value $\Omega_c$ than the actual initially measured $\Omega_c^*$
at the sudden jump. This is an example where the global order
parameter texture becomes unstable in the increasing superflow and
finally a first order transition occurs in the texture to a new
configuration with a lower critical velocity. It is also clear
evidence for the fact that the value of the critical velocity depends
on the global order parameter texture in the rotating container.

Measurements \cite{RuutuA} and calculations \cite{Kopu} of the
critical velocity in $^3$He-A show that the maximum limit for the
critical velocity is reached with an ordered texture
which mimics one where $\hat{\bf l}$ is homogeneously oriented
along the superflow ${\bf v}_s$. The minimum velocity, in turn, is
close to an order of magnitude smaller and is obtained within a
soliton sheet where the spin-orbit coupling is broken and the
$\hat{\bf l}$ texture is inhomogeneous.

\subsection{Formation of vortex sheet}

The dependence of the critical velocity on the core size explains why
vorticity with continuous singularity-free structure is formed, when
$^3$He-A is accelerated to rotation. But is it created in the form of
lines or sheets? This has turned out to be an interesting question of
general validity. The vortex sheet is formed whenever a vertical
dipole-unlocked soliton sheet is present in the container, while
rotation is started. Here the critical velocity is the lowest
possible, ie. the energy barrier for adding more merons into the sheet
vanishes at lower velocity than for any other type of vortex
structure.

The critical velocity of the vortex sheet is made up of several
contributions. First of all there is the low critical velocity at the
connection lines between the sheet and the cylinder wall, where the
spin-orbit coupling is broken.  Some low superflow velocity is
required even here,  since at least the attractive interaction of the
emerging meron with its image within the wall has to be overcome. A
second contribution is a small, but nonzero resistance of the texture
in the soliton connection line to reach the instability limit. These
small contributions at the connection line are the only ones which are
effective initially when rotation is started and the first circulation
quanta enter the sheet.

\begin{figure}[t!!!]
\centerline{\includegraphics[width=0.85\textwidth]{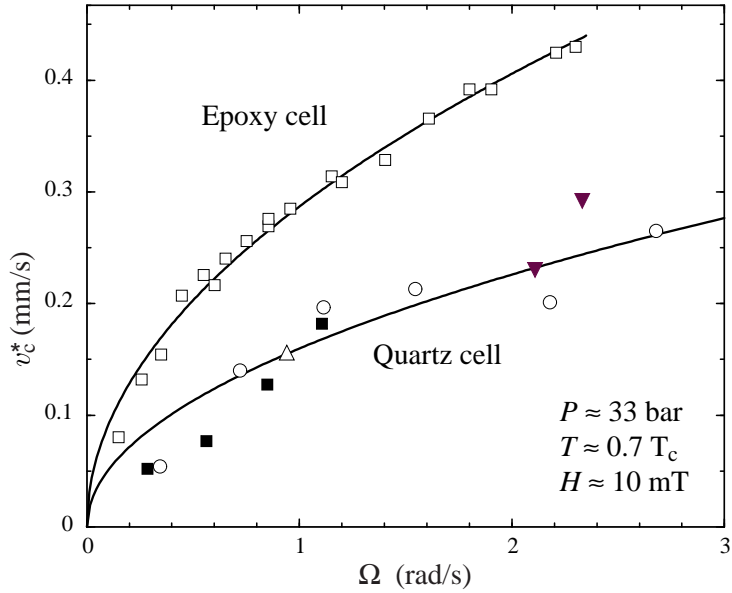}}
\caption[CritVel-VorSheet]{ Critical velocity $v_c^*$ of the
adiabatically grown vortex sheet as a function of applied rotation
$\Omega$: This state consists of a single sheet, which in an axially
oriented magnetic field is folded into a double spiral, as shown in
Fig.~\protect\ref{VorSheet}. The solid curves are fits to the measured
data with the dependence $v_c^* \propto \sqrt{\Omega}$. An epoxy
container with surface roughness of order $10\,\mu$m displays a larger
magnitude of $v_c^*$ than a container with fused quartz walls and an
order of magnitude smaller roughness ($\lesssim 1\,\mu$m). The
difference could arise from slight pinning of the two connection lines
of the sheet along the cylindrical wall which would resist
readjustments when new vorticity is added into the sheet during
acceleration. The different symbols of data points (quartz cylinder)
illustrate the reproducibility of the results from one adiabatically
grown vortex sheet to another. } \label{CritVel-VorSheet}
\end{figure}

As the sheet grows, a second contribution becomes effective. When a new
meron is added, it experiences repulsion from the meron which already
resides in the sheet close to the connection line. The repulsion depends on
the distance between the merons in the sheet and the sheet's resistance to
change its folding and the distribution of circulation in the container.
Thus the critical velocity becomes $\Omega$-dependent. At low $\Omega$,
when the merons are rare, this contribution is much below the critical
velocity for the formation of isolated vortex lines or skyrmions. It was
suggested and later experimentally confirmed that the critical velocity
follows the qualitative dependence $v^*_c(\Omega) \propto \sqrt{\Omega}$ as
shown in Fig.~\ref{CritVel-VorSheet}. (The star as a superscript marks the
fact that $v^*_c$ is obtained from the measured value of $\Omega_c$ through
the relation $v^*_c = \Omega_c/R$ which is strictly valid only for an
axially symmetric distribution of vorticity.)

The vortex sheet has unusual dynamic properties which make it to the
preferred state if the rotation drive is rapidly changing in time
\cite{Eltsov2000}. It is expected that analogous features can evolve
in anisotropic p-wave superconductors with periodic vortex flow in the
unpinned regime \cite{Kita}. The dynamic response of $^3$He-A as a
function of the frequency and amplitude of the external drive is
currently under study \cite{Eltsov2001}. The time proven method to
create the vortex sheet is to apply sinusoidally modulated rotation in
the form $\Omega = \Omega_1 \sin{\omega t}$, where the period of
modulation is of order $2\pi/\omega \sim 10\,$s and the amplitude
$\Omega_1 \sim \hbar/(2m_3\xi_D R) \sim 0.3\,$rad/s exceeds that
required for breaking the spin-orbit coupling. To grow the equilibrium
vortex sheet, one applies the oscillating rotation until the signal
from the vortex sheet is observed \cite{Sheet}. The final step is to
increase $\Omega$ from zero to the desired value, using slow
acceleration $(d\Omega/dt \lesssim 10^{-3}$ rad/s$^2$) to ensure
adiabatic growth of a single folded sheet.

\subsection{Vortex formation in ionizing radiation}

In the metastable regime of superflow at $v_s <v_c$, vortex formation
can be triggered by irradiation with ionizing radiation. Experiments
with superfluid $^3$He-B \cite{BigBang} have shown that quantized
vortex lines are formed in the aftermath of a neutron absorption
event. According to current belief, vortex formation occurs via the
Kibble-Zurek (KZ) mechanism \cite{Kibble,Zurek}, which was originally
developed to explain the phase transitions of the Early Universe. In
this scenario a network of cosmic strings is formed during a rapid
non-equilibrium second order phase transition, owing to thermal
fluctuations. In $^3$He the absorption of a thermal neutron causes
heating which drives the temperature in a small volume of
$\sim100\,\mu$m size above the superfluid transition. Subsequently the
heated bubble cools back below $T_c$ with a thermal relaxation time of
order 1\,$\mu$s. This process forms the necessary conditions for the
Kibble-Zurek mechanism within the cooling bubble. It is interesting to
note that so far this is the only case of vortex formation in the
$^3$He superfluids which in principle is not confined to the vicinity
of the bounding solid walls. In practice the mean free path of thermal
neutrons in liquid $^3$He is only $100\,\mu$m and therefore even this
process is localized close to the wall.

The real experimental conditions in the neutron irradiation experiment
of $^3$He-B \cite{Review} (and also probably in the early Universe) do
not coincide, with a perfectly homogeneous transition, as is assumed
in the KZ scenario: The temperature distribution within the cooling
``neutron bubble'' is nonuniform, the transition propagates as a phase
front between the normal and superfluid phases, and the phase is fixed
outside the bubble. These considerations require modifications to the
original KZ scenario \cite{KV,DLZ,KT} and even raise concerns whether
the KZ mechanism is responsible for the defects which are extracted
from the neutron bubble and observed in the experiment
\cite{Ruutu2,AKV}. New measurements demonstrate that a joint defect --
the combination of a conventional (mass) vortex and spin vortex -- is
also formed and is directly observed in the neutron irradiation
experiment \cite{CompositeDefect}. This strengthens the importance of
the KZ mechanism and places further constraints on the interplay
between it and other competing effects.

In superconductors localized heating can cause the unpinning of
vortices from defects in the crystal lattice which can be viewed as a
creation of an intermediate vortex ring.

\section{Vortex dynamics without pinning}

As a liquid free of alien impurities, He superfluids do not display
bulk pinning. Surface pinning at protuberances on the container wall
remains an issue which has been studied in $^4$He-II \cite{Glaberson}.
In $^3$He superfluids vortex core diameters are at least two orders of
magnitude larger and pinning correspondingly weaker. Measurements of
vortex dynamics have so far not resulted in reliable estimates for
pinning parameters. In fact, it has not been settled whether surface
pinning plays any observable role in the presence of smooth walls. All
indications point in the direction that, even in $^3$He-B with its
smaller vortex core sizes, vortex motion occurs in the limit of weak
pinning \cite{Sonin}, where collective effects are expected to
dominate in pinning. For vortex lines with continuous structure in
$^3$He-A pinning is expected to be unimportant. The virtual absence of
pinning has a number of important consequences:

(i) It allows investigation of vortex dynamics without pinning in the
whole temperature range from $T=0$ to $T=T_c$. As a result, in
$^3$He-B three topologically different contributions to vortex
dynamics have been distinguished from one another, owing to their
different temperature behavior \cite{BevanJLTP}. These are 1) the
Magnus force with which the flowing superfluid component acts on a
vortex line, 2) the Kopnin force which is caused by the spectral flow
phenomenon \cite{KopninSpectralFlow} and which acts on the vortex line
when it moves with respect to the normal component, and 3) the
Iordanskii force which is the analog of the gravitational
Aharonov-Bohm effect \cite{StoneIordanskii}.

(ii) It becomes possible to prepare surfaces with specially prepared
pinning sites by micro-fabrication techniques. One can then study
vortex dynamics when an array of vortex lines becomes commensurate
with a pre-fabricated lattice of surface pinning sites.

(iii) One can study the trapping and unpinning of circulation from a
columnar defect in the order-parameter field -- namely a thin wire
stretched across the superfluid bath parallel to the rotation axis.
This is the Vinen vibrating wire configuration where one quantum of
circulation can be trapped around the wire by rotating the container
\cite{VinenWire}. When rotation is stopped, the trapped circulation
can be observed to peel off from the wire as a precessing vortex
\cite{Packard}. With each revolution of the precessing vortex, the
phase difference between the ends of the container slips by $2\pi$.
This process can be interpreted as a macroscopic manifestation of the
ac Josephson effect \cite{Misirpashaev,PackardRMP}. The corresponding
Josephson frequency is remarkably low, approximately 4\,mHz.

(iv) Unimpeded by pinning on solid walls, it becomes possible to study
interactions of vortex lines with different types of interfaces which
can be prepared in the He systems. These interfaces include the free
surface of the superfluid bath (i.e. the gas -- liquid interface)
\cite{Manninen}, the superfluid -- solid $^3$He interface
\cite{Parshin}, the interfaces between $^3$He and $^4$He superfluids
\cite{Sonin,Korhonen}, and interfaces between the A and B phases in
superfluid $^3$He \cite{ABinterface}.

\section{Conclusion}

Although superflow and quantized vortex lines have been the essence of
superfluid investigations since the start in the late nineteen
forties, nevertheless uniform rotation has not become an important
tool for generating vorticity in $^4$He-II, owing to the uncontrolled
release of remanent vorticity. When the construction work of the first
rotating nuclear demagnetization cryostat was started in 1978 it was
feared that rotation might not prove a useful concept in $^3$He
superfluidity either. However, today we know that the $\Omega$ axis is
as important in the study of the $^3$He superfluids as the other
experimental parameters $H$, $P$, or $T$, which together control the
various order-parameter structures: Measurements on $^3$He superfluids
without access to rotation would be as limited as studies on
superconductors without the possibility to turn on the magnetic field!

The central areas so far in superfluid $^3$He work have been the
identification of different topologically stable defect structures in
the order-parameter field, the conditions and critical values which
control their formation, and phase transitions between different
structures as function of the external variables. For these questions
the $^3$He superfluids have been the most ideal system, owing to the
large variability in its order-parameter response. Today the largest
collection of different theoretically characterized and experimentally
identified quantized vortex structures are found in the $^3$He
superfluids. In the coming years $^3$He work will focus more and more
on making use of its ideal text-book-like properties, to employ the
$^3$He liquids as a model system in quantum field theory for the study
of such varied questions as the physical vacuum, black hole, or
inhomogeneity in the accelerating Universe. For many questions of this
category working theoretical $^3$He analogues have already been
constructed \cite{VolovikReview}.

%

\end{document}